\documentclass[runningheads]{llncs}
\usepackage{graphicx,xcolor,amsfonts,booktabs}

\begin{document}
\titlerunning{A-Muze-Net}
\title{A-Muze-Net: Music Generation by Composing the Harmony based on the Generated Melody}

\author{
Or Goren\inst{1}
\and
Eliya Nachmani\inst{1,2}
\and
Lior Wolf\inst{1}
}
\authorrunning{}

\institute{$^1$The Blavatnik School of Computer Science, Tel-Aviv University\\ $^2$Facebook AI Research}
\maketitle              % typeset the header of the contribution
\begin{abstract}
We present a method for the generation of Midi files of piano music. The method models the right and left hands using two networks, where the left hand is conditioned on the right hand. This way, the melody is generated before the harmony. The Midi is represented in a way that is invariant to the musical scale, and the melody is represented, for the purpose of conditioning the harmony, by the content of each bar, viewed as a chord. Finally, notes are added randomly, based on this chord representation, in order to enrich the generated audio. Our experiments show a significant improvement over the state of the art for training on such datasets, and demonstrate the contribution of each of the novel components.

\end{abstract}
\section{Introduction}

We present a new method of symbolic music generation called A-Muze-Net. The method employs relatively low-capacity models, such as LSTM networks, and is trained on a relatively small dataset. In order to generalize well despite the lack of training data, it employs various techniques that are inspired by the common practices of human composers.

First, the harmony is composed after the melody is determined, and is conditioned on the melody. Second, the notes are represented in a way that is scale invariant, by considering the gap in pitch between the notes. Third, instead of separating notes to pitch and length, a single token captures both.

A crucial component of the method is that the melody is encoded by considering the notes at each generated bar and identifying the closest chord to this set of notes. Finally, a heuristic that employs the same chord-view adds random notes in order to make the generated audio more complete. 

We demonstrate the advantage of our method over existing methods using a collection of 243 Midi files of Bach music. In addition, an ablation study demonstrates the value of each of the above mentioned contributions.

\section{Related Work}

Music generation methods can be divided into a few categories based on the generation domain. Many of the recent works, generate raw audio. \textbf{WaveNet}~\cite{oord2016wavenet} is a convolutional neural network which inputs are raw audio files, and it generates new raw audio files. Since each raw audio timestamp is represented as a 16-bit integer, quantization is applied to reduce the output space~\cite{recommendation1988pulse}. A followup research by Manzelli et al.~\cite{manzelli2018end} employs the same quantization method but employs a biaxial LSTM model~\cite{johnson2017generating} which is built as two different LSTM models, one for the time-axis and one for the note-axis. The note-axis LSTM takes two inputs, the previous input as well as the final output from the time-axis LSTM model. \textbf{MP3net}~\cite{broek2021mp3net} generates new mp3 files given mp3 files by using a CNN, their representation is based on the mp3 compression. The most evolved out of these methods is \textbf{Jukebox}~\cite{dhariwal2020jukebox}, which compresses the raw audio data to discrete representation and apply high capacity transformer networks~\cite{vaswani2017attention} to generate songs from many music genres, such as rock and jazz. It is trained on massive amounts of recorded data.

Many of the classical music composition approaches generate music scores~\cite{yang2017midinet,papadopoulos2016assisted,yan2018part}, and this line of work has continued into the era of deep learning. \textbf{MidiNet} ~\cite{yang2017midinet} employs a GAN in which both the generator and the discriminator are CNN models, \textbf{FlowComposer}~\cite{papadopoulos2016assisted} which uses Constrained Markov Models~\cite{roweis1999constrained}. The Part-Invariant model~\cite{yan2018part} is a single RNN layer model that generates a composition based on an initial part.

\begin{figure}[t]
\includegraphics[width=\textwidth]{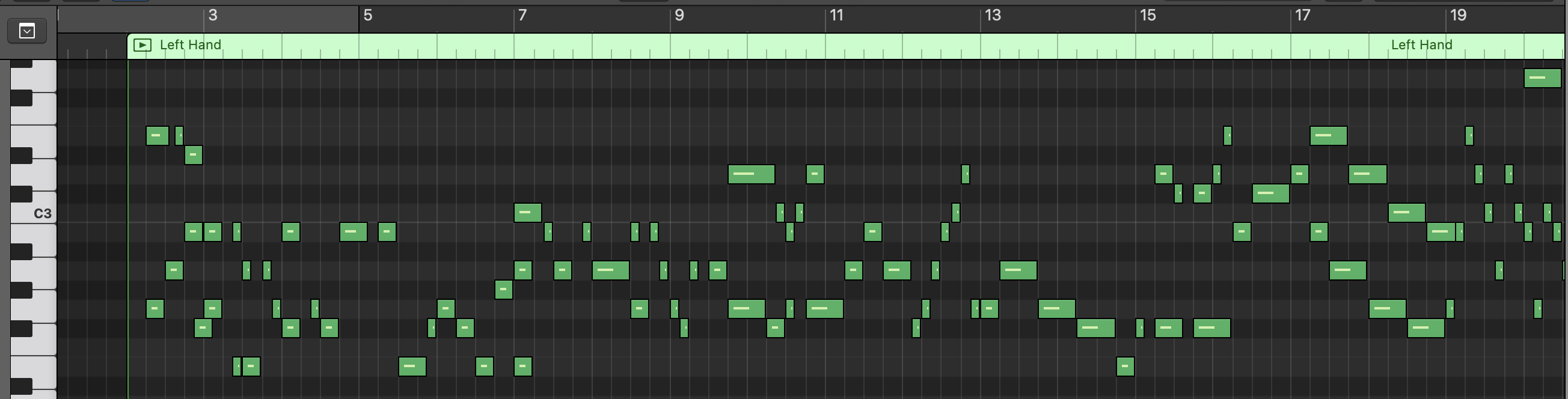}
\caption{Midi representation of prefug3 left-hand track, taken from Complete Bach Midi Index Dataset and opened with Logic Pro software} \label{fig_midi}
\end{figure}

Our model generates Midi notes given a prompt that it continues. Recent works that perform the same task include \textbf{MuseGan}~\cite{dong2018musegan} by Dong et al., which generates novel multi-track Midi files using a GAN model~\cite{goodfellow2014generative} trained on a large scale dataset. A multi-track Midi file contains a separate track for multiple instruments, such as guitar, piano, and drums. It is represented as a \textbf{Multi-Track Piano-Roll}. A single Piano-Roll is illustrated in Fig.~\ref{fig_midi}, which is a binary-valued matrix where each row index represent a pitch value and each column index represent a time frame. 
Dong et al. have later on used a convolutional GAN to generate polyphonic music~\cite{dong2018convolutional}. Binary neurons are used to generate the binary piano-roll representation, which was found to be more successful than using regular Hard Threshold (HT) or Bernoulli Sampling (BS) as was done in their earlier work.

Boulanger-Lewandowski et al.~\cite{boulanger2012modeling} also used the piano-roll representation but employed the Restricted Boltzmann Machine (RBM) on top of the RNN model in order to generate high-dimensional sequence. 
The dual-track generator of Lyu et al.~\cite{lyu2020dual} generates piano classical music. Similar to our method, it first generates the right-hand part and then the left-hand. In their model, the right-hand is generated by an LSTM, and the left-hand is subsequently generated using a Multi Layer Perceptron. Our left-hand generator is considerably more evolved as it's an LSTM that considers the chord embedding of the right-hand. In addition, while Lyu et al. represent the data as Piano-Rolls, we employ the normalized Midi representation, similar to \textbf{BachProp}~\cite{colombo2018bachprop}.

\textbf{DeepJ}~\cite{mao2018deepj} generates specific music styles: Baroque, Classical and Romantic. Similar to MuseGAN, a piano roll is employed as the underlying Midi representation. They employ a biaxial LSTM model~\cite{johnson2017generating}, similar to the approach of~\cite{manzelli2018end}. Two LSTM models are used, one for the notes pitches and one for the duration of each such note. In order to pass the music genre information an embedding layer is used.

\textbf{Music Transformer}~\cite{huang2018music} is a transformer model with relative attention that generates symbolic music based on multi-track piano-roll representation of Bach's Chorales dataset. For the Piano-e-Competition dataset they have used the Midi events as their domain.
\textbf{BachProp} is another LSTM model which is trained on given Midi files and composes new compositions. The normalized Midi representation it employs transforms the Midi into a sequence of notes, each with an associated length. 
The representation is defined as follows $note[n] = (T[n], P[n],dT[n])$, where $T[n]$ is the duration of the note, $P[n]$ is the pitch and $dT[n]$ is the time interval between the current note and the previous one. Their implementation employs three LSTM models, one per each input.
\textbf{DeepBach}~\cite{hadjeres2017deepbach} is a deep learning model that uses the Bach's Chorales dataset, and generates new Chorales like Midi files. They are using only the Chorales, separating these to four different voices, where each voice has a single note at a time.
Wu et al.~\cite{wu2019hierarchical} employed a Hierarchical-RNN (HRNN) to generate symbolic music. This was done using a slightly different Midi representation, in which the input domain is the Midi events note-on and note-off, and the time interval since the last event, like~\cite{huang2018music}. Their HRNN is built from three conditioned RNN models based on the bars, beat and notes. 

Our representation is slightly different and employs notes and duration only. Furthermore, we employ a scale-invariant representation, see Sec.~\ref{sec:method}. In addition, we employ one LSTM for each hand, which are trained subsequently, and do not split the LSTM networks by the type of information of the note tuple. 

\section{Method}
\label{sec:method}

We describe the way the midi file is represented, the model we propose, and its training.

\subsection{Midi representation}

Given a Midi file, we apply a parser that outputs the note's pitch and length values for each note in the Midi file for each track. The obtained representation follows closely the music notes representation. The output is a string, in which the alphabet is a sequence of tokens. A sample token is `5-X-1/8', which means note `X' at the fifth octave and length 1/8. 

\begin{figure}[t]
\begin{center}
\includegraphics[width=.607\textwidth]{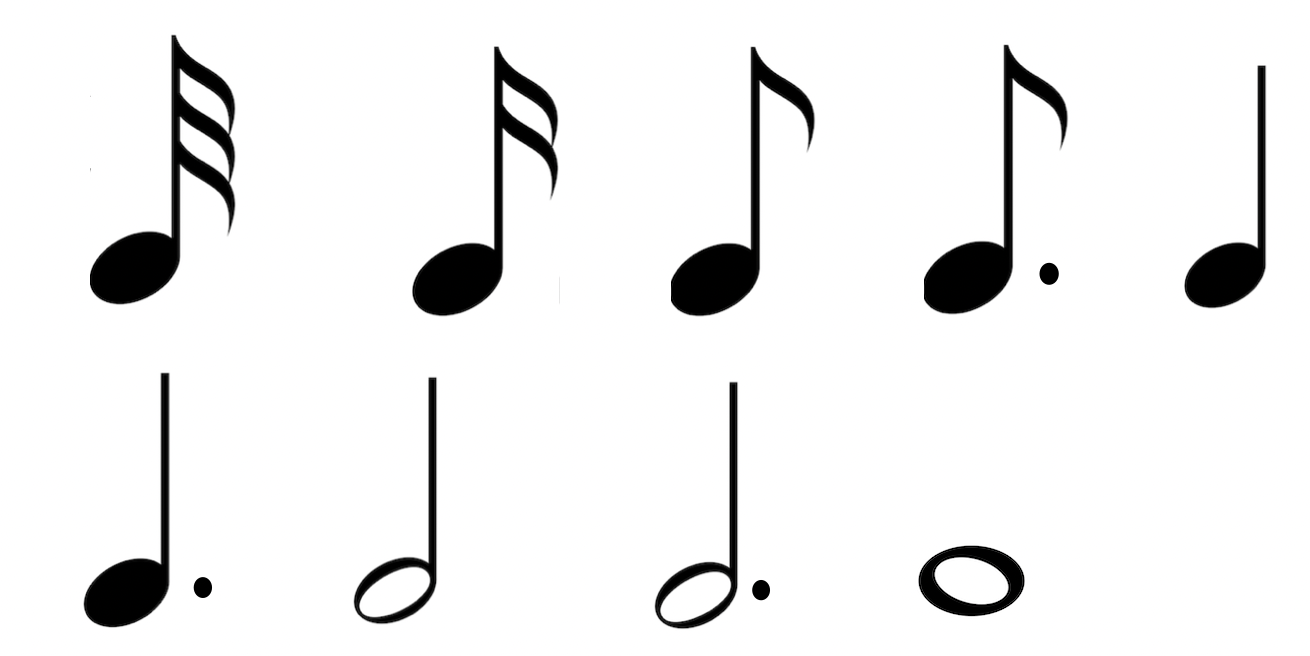}
\end{center}
\caption{All allowed notes lengths, First line from left to right: 1/32, 1/16, 1/8, 3/16 and 1/4. Second line from left to right: 3/8, 1/2, 3/4 and 1. The note images were obtained from \url{https://www.freepik.com/}. } \label{fig_notes}
\end{figure}

Each token is converted to an integer in the following way. First, we quantize the note's length into one of the following common values: {1/32, 1/16, 1/8, 3/16, 1/4, 3/8, 1/2, 3/4 and 1}, as illustrated in Fig.~\ref{fig_notes}. If, for example, the note's length is $11/16$ which is rare, then we assign it to be $3/4$. 

The note integer representation which the networks employ is the product space of the nine length values and the 128 possible values of a note's midi-num\footnote{The Midi-num table assigns for each piano keyboard note a number, see \url{https://computermusicresource.com/midikeys.html}}.

The notes themselves are not taken as absolute notes, such as C (do), D (re), etc. Instead, we represent the Midi data in a way that is invariant to the musical scale used. A Midi file contains the scale information, and we compute each note's interval in the scale from the first note of the scale.

Two separate sequences are then generated. Specifically, for piano music one sequence is generated for the right-hand and one for the left. In some of the Midi files of the dataset, the separation is not provided. To overcome this, the average pitch for each track is calculated, and the track which has the lowest average pitch is chosen to be the left-hand track, and the maximal average pitch track is chosen to be the right-hand track. This stems from the position of the left hand on the keyboard relatively to the right hand, on the side of the lower notes.

Our representation assumes that there is no more than one note played simultaneously for each track. In case that the input contains multiple simultaneous notes, the right-hand selects the note with the highest pitch, and the left-hand the one with the lowest. Thus, heuristic relies on the observation that in the melody (right-hand) the highest note is more descriptive than the other ones, and vice versa for the harmony (left-hand).

\subsection{Models}
\label{sec:models}

Our method relies on the common practice to compose the melody first and the harmony afterwards\footnote{See, for example, \url{https://www.artofcomposing.com/how-to-compose-music-part-3-melody-or-harmony-first}}. Therefore, the \textbf{Right Hand LSTM Model} is applied first. Subsequently, the \textbf{Left Hand LSTM Model} is applied conditioned on the output of the \textbf{Right Hand LSTM Model}. 

\begin{figure}
\includegraphics[width=\textwidth]{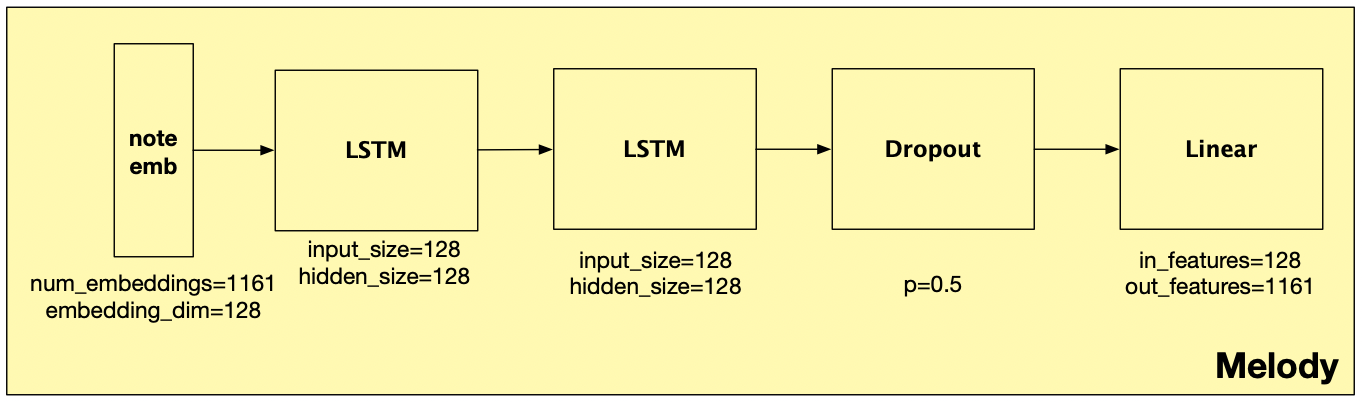}
\caption{Right Hand LSTM Model Architecture} \label{fig_melody}
\end{figure}

The architecture of the \textbf{Right Hand LSTM Model} is depicted in Fig.~\ref{fig_melody}. The embedding layer converts each of the possible $1161$ integer values of the music representation into an embedding vector in $\mathbb{R}^{128}$. This is followed by two LSTM layers with a hidden size of the same dimensionality (128) and there are two layers. This is followed by a dropout layer with a factor of 0.5. Finally, a linear layer projects the LSTM output to a vector of length 1161 that produces the pseudo-probabilities (using softmax) of the next element in the sequence. 

In our method, the harmony generator network is conditioned on a new type of signal we propose, which is the chord analysis of the melody.

\begin{figure}
\includegraphics[width=\textwidth]{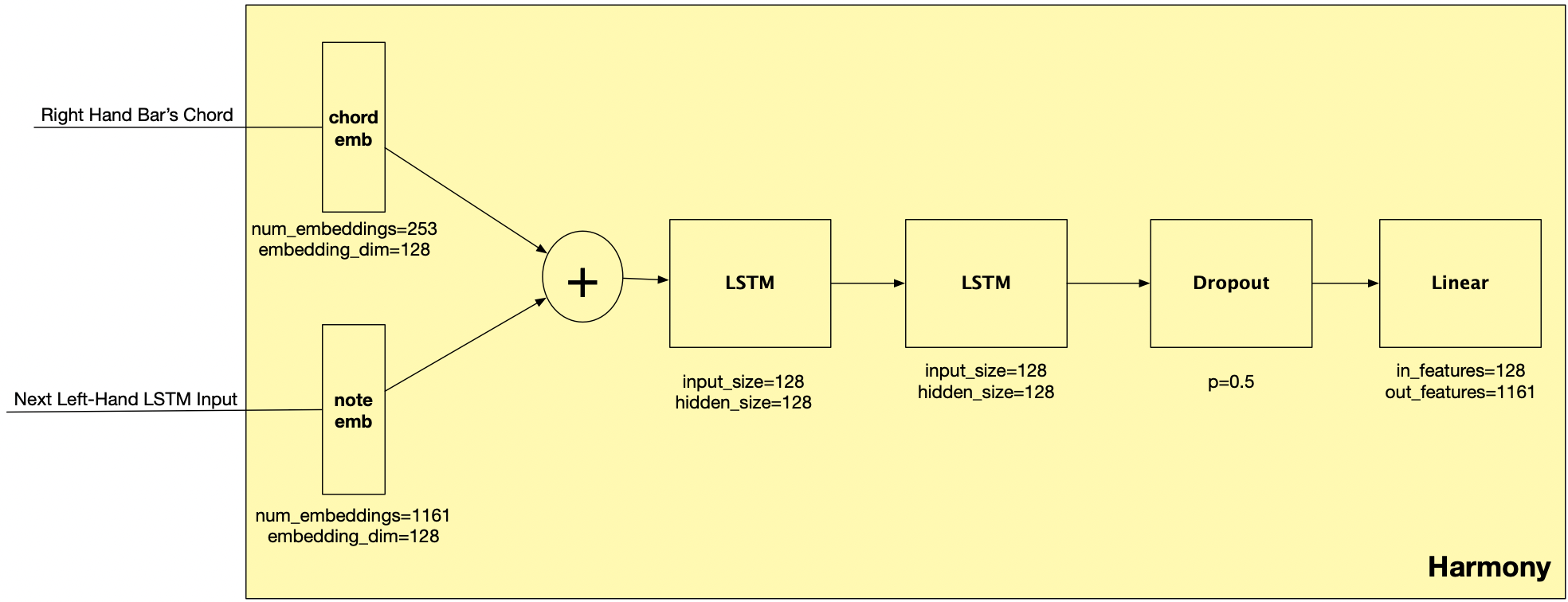}
\caption{Left Hand LSTM Model Architecture} \label{fig_harmony}
\end{figure}

The architecture of the \textbf{Left Hand LSTM Model} is depicted at Fig.~\ref{fig_harmony}. It is similar to the \textbf{Right Hand LSTM Model}. The main difference is that a second embedding layer is used. This added embedding, termed the \textbf{Chord Embedding Layer} captures the chord that is being played by the right-hand on the current bar. The number of items this embedding encodes is 253 and the embedding dimension is 128. Each of the 253 options encoded a specific combination of notes in the current bar played by the Right Hand Side. Formally, the input of the LSTM after the embedding is:
\begin{equation}
    x_{emb} = E_{notes}(x) + E_{chords}(f(R))
\end{equation}
where $x_{emb}$ is the LSTM input, $x$ is the current left-hand note, $E_{notes}$ is the \textbf{Notes Embedding Layer}, $E_{chords}$ is the \textbf{Chords Embedding Layer}, $R$ is the current right-hand bar's notes and $f$ is a function that maps a list of notes to their correspondent chord.

For this purpose, we employ a chords hash table that maps the chord's name to its right form, for example the chord \textbf{C} is mapped to notes \textbf{"C","E","G"}. Also we gather groups of 7-chords, 6-chords, 13-chords, 9-chords, diminished chords and augmented chords, by adding/changing the original chord values. For example the chord \textbf{Cmaj7} is constructed by \textbf{"C", "E", "G", "B"}, and \textbf{Ddim} is constructed by \textbf{"D", "F", "$G^\#$"}. We apply this method to all possible chords and finally obtain the 253 chords mentioned above.

For recovering the chord associated with a specific bar, we gather all of the notes in that bar, and each chord is scored based on how many notes from a given bar belong to this chord.  For example, if the notes are \textbf{"C", "E" and "G"} then chord \textbf{Cmaj7} will get a score of 3 and chord \textbf{Am} will get a score of 2, However chord \textbf{C} will get a score of 13 as these notes are the exact notes within the \textbf{C} chord, and it would be picked up. In case of a tie the more common  chord would be chosen, i.e., \textbf{D} would be picked up rather than \textbf{D7}.

\subsection{A-Muze-Net Model Training}

\begin{figure}
\includegraphics[width=\textwidth]{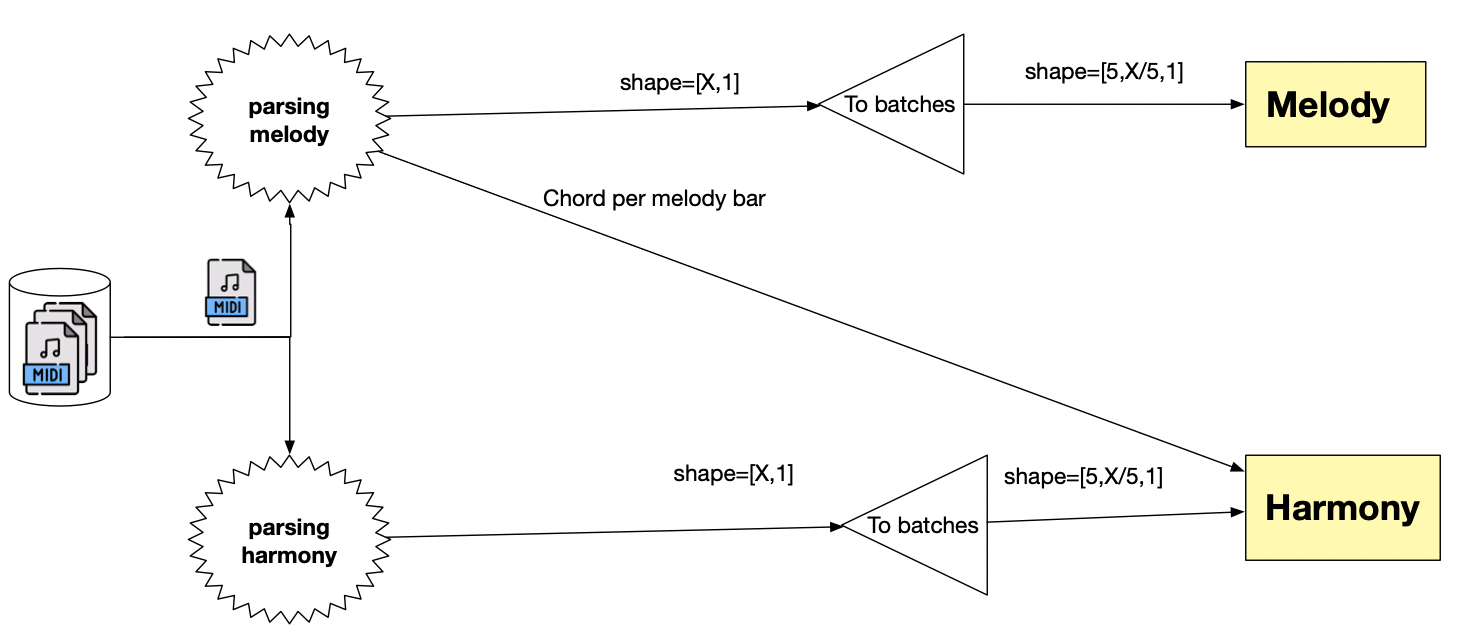}
\caption{A-Muze-Net Model Architecture} \label{fig_arch}
\end{figure}

The model training process is depicted in Fig.~\ref{fig_arch}. The given dataset is constructed from multiple Midi files, and we feed one Midi file at a time to the A-Muze-Net Model. The Midi file goes through the parsing methods, and is then divided into batches (the batches do not mix between multiple files).

\subsubsection{Prediction} -
The prediction method inputs are:
\begin{enumerate}
    \item A list of initial Melody notes (right-hand)
    \item A list of initial Harmony notes (left-hand)
    \item The number of notes to generate
\end{enumerate}
At start the hidden layers of the two LSTM models are initialized and are not being reset throughout the generation part, in order to keep the composition context. 
The initial melody sequence is fed to the already trained melody LSTM model while using teacher forcing to preserve the specified initial melody. This will set the hidden layer to preserve the initial melody context. Then, the last output from the initial feeding is essentially the first note prediction by the LSTM model. As the output is a probability for each note, only the top-k values are considered and a random value is chosen amongst them, after filtering the "break" notes. Then the chosen note is the next input of the melody LSTM model and again chooses the next predicted note from the next top-k ones. This process is finished when the given \textbf{number of notes to generate} is reached.

After this process is finished the chords for each melody bar are calculated and preserved as a sequence for chord per bar. 

Then, the Harmony LSTM (Left-hand) is trained the same way as the other one, but the current right-hand bar's chord is passed through to the \textbf{Chord Embedding Layer} while the notes are passed through to the \textbf{Note Embedding Layer}. The initial sequence is also done in this method with teacher forcing, while the rest of the sequence up until the \textbf{number of notes to generate} is done without any teacher forcing.

After both the right-hand track and the left-hand track are generated, we apply a heuristic to add simultaneous notes. For each note in the harmony which is inside the current composition's scale, there is a 50\% chance that it'll be accompanied with its perfect-fifth interval note and another 50\% chance that it'll be also accompanied by its third interval note which is in the scale. 

For example if the current scale is \textbf{C} and the current harmony note is \textbf{E} then in 50\% chance the note \textbf{B} (fifth) would be added simultaneously, and another 50\% that \textbf{G} (third - to form \textbf{Em} essentially) would be added.

For out of scale notes the method is different. Instead of adding their fifth or third interval, we add this note again but from a different octave. 

For example, if in a composition of scale C the current note is $4-B^b$, for which $B^b$ is at the fourth octave, then there is a 50\% chance that $5-B^b$ would be added as well. 

For harmony, we apply a similar heuristic. However, instead of a 50\% chance of generating simultaneous notes, there is a 10\% to generate them since the harmony often generates simultaneous notes to form chords.
\section{Experiments}

For our experiments, we employ the \textbf{Complete Bach Midi Index} dataset\footnote{\url{http://www.bachcentral.com/midiindexcomplete.html}}. It contains approximately 243 Midi files which are divided into several different genre topics, such as chorales, cantatas, fugues and more. The Midi files have different scales and different time-signatures. Most of these Midi files contain two tracks, one for the right-hand (melody) and one for the left-hand (harmony).

As baselines we employ two methods: \textbf{MuseGan}~\cite{dong2018musegan} and to Lyu et al. research~\cite{lyu2020dual}. Since our model is piano-based, we compare our results with the MuseGan piano track.

In addition to comparing with the baselines, we also compare with two ablation variants of our model. 

\paragraph{Ablation 1 - no conditioning of the harmony on the melody -}
\label{sec:ablation1}
In this experiment we trained the right-hand (melody) LSTM model and left-hand (harmony) LSTM model separately, meaning that we canceled the \textbf{Chords Embedding Layer} from the left-hand LSTM model. This way, both models are independent on one another. After the training was done we generated the Midi files.

\paragraph{Ablation 2 - note embedding instead of chord embedding -}
\label{sec:ablation2}
In this ablation, we maintain the conditioning of the harmony on the melody, but instead of the \textbf{Chords Embedding Layer} we simply take the summation of all of the notes' embeddings of the right-hand bar and add them to the current left-hand note's embedding. In other words, we employ the following embedding 

\begin{equation}
    x_{emb} = E_{notes}(x) + \sum_{n\in R}E_{notes}(n)
\end{equation}
Where $x$ is the current left-track note, $x_{emb}$ is the embedding output and the input to the LSTM layer, $E_{notes}$ is the \textbf{Notes Embedding Layer}, $R$ is the list of the corespondent right-hand bar notes.

\paragraph{Ablation 3 - without the notes addition method -}
\label{sec:ablation3}
In this ablation, we remove the part that adds random harmonic notes. In other words, the results are the output of the LSTM models without any further post-processing steps

\begin{figure}[t]
\centering
\begin{tabular}{cc}
\includegraphics[width=.495\textwidth]{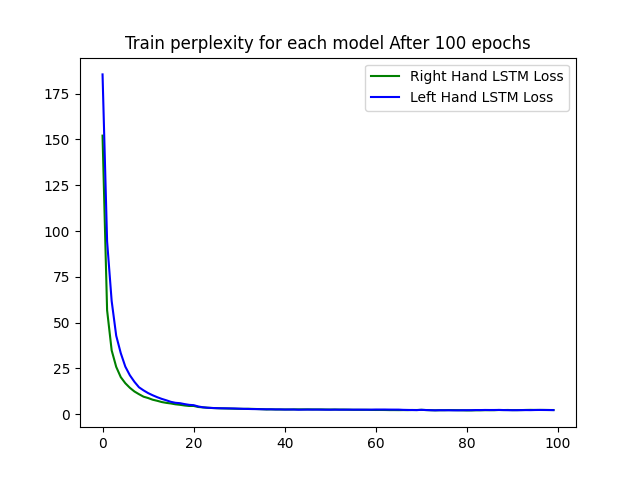}&
\includegraphics[width=.495\textwidth]{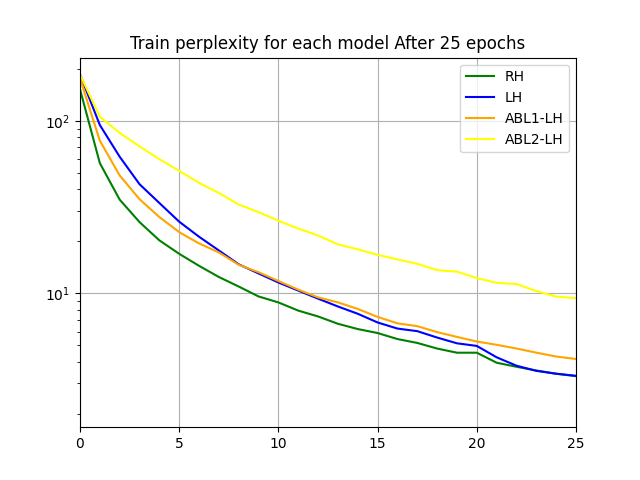}\\
(a) & (b)\\
\end{tabular}
\caption{(a) Train Perplexity For the Right-Hand LSTM Model and for the Left-Hand LSTM Model. (b) Train Perplexity with Ablation Studies.} \label{fig_train_abl} \label{fig_train_perp}
\end{figure}

The training perplexity of the full method is shown in Fig.~\ref{fig_train_perp}(a). As can be seen, while the harmony and melody seemingly converged kind of the same way, still the melody converged faster than the harmony.

The training perplexity of the full method compared to the ablation studies training perplexity is shown in Fig.~\ref{fig_train_abl}(b). One can observe that \textbf{ABL1-LH}, which is the Left-Hand LSTM model of ablation study 1 is slightly above our Right-Hand LSTM model and below our Left-Hand LSTM model. This is expected since the \textbf{ABL1-LH} is not dependant on the Right-Hand LSTM model, and so its perplexity should be the same as the Right-Hand LSTM model. As for \textbf{ABL2-LH}, the obtained loss is considerably above our Left-Hand LSTM model and it takes it longer to converge. This emphasizes the importance of the \textbf{Chords Embedding Layer}.

\subsection{Results}
\label{sec:Results}
Following previous work, we consider the following quantitative metrics:
\begin{enumerate}
    \item \textbf{QN (Qualified Note)} - The percentage of notes that were generated with a valid length. For example a note with length lower than 1/32 is considered faulty.
    \item \textbf{UPC (Unique Pitch Class)} - The average amount of different pitches per bar.
    \item \textbf{TD (Tonal Distance)} - A number that specifies how much the two tracks are aligned chromatically, lower numbers are better.
    \item \textbf{OOS (Out Of Scale)} - The percentage of generated notes that were out of scale.
\end{enumerate}

Tab.~\ref{tab1} presents the results for our algorithm in comparison to the baselines. As can be seen, our model's \textbf{QN} achieves 100\%, which means that all notes have a valid length size. This is because we maintain an allowed lengths list that each generated note has one of these lengths. This way, all of the generated notes are of qualified note lengths, and our method does not get fragments of notes, e.g. notes with length less than 1/32. Also we can see that our model achieves the closest \textbf{UPC} value to the True Music UPC value. For the \textbf{TD}, where lower values are better, we achieved a lower TD value than \textbf{MuseGan}, which means that our tracks are more coherent to each other. 

Tab.~\ref{tab1} also depicts the results of the ablation study. Evidently,  the UPC values for the ablation methods are lower, which indicates that the tracks are not aligned and do not complete one another. The second ablation study achieved the lowest score which might indicate that the addition of many right-hand notes together with the current left-hand one maybe interfere with the learning method of the model, causing it to generate much fewer notes. We can also observe that the TD values are much higher, and as expected the TD value of the first ablation study is higher than the TD value of the second one, which means that the model with no conditioning at all achieved a worse coherence score. 

{
Our model has a higher OOS percentage in comparison to our ablation studies, which is consistent with the model's higher UPC value. In a music scale there are only seven notes which are inside the scale, and the five others are considered as out of scale. Since we have UPC value which is higher than seven,  we have a high percentage of out of scale notes. Almost all music compositions uses notes from out of scale to generate unique sounds, as is evidenced from the high UPC value of True Music baseline.  For example if a composition at  scale C uses the \textbf{D} chord, it necessarily uses the $F^\#$ note which is out of scale. Interestingly, the second ablation study achieved 20\% OOS although it uses less than seven notes, which reveals a mismatch {with} the notes {being} used { at the harmony side}, pointing to the significance of using the Chord Embedding Layer. 
}

\begin{table}
\begin{center}
\begin{tabular}{lccccc}
\toprule
\textbf{Experiment} & \textbf{QN} & \textbf{UPC} & \textbf{TD} & \textbf{OOS}\\
\midrule
True Music & 98.70\% & 9.83 &  - & - \\
Lyu et al. Pianoroll CNN~\cite{lyu2020dual} & 91.20\%  & 2.35 & - & -\\
Lyu et al. Embedding Atten-LSTM~\cite{lyu2020dual} & 90\% & 7.79 & - & -\\
MuseGan & 64\% & 4.57 & 0.94 & -\\
\midrule
A-muse-Net - Ablation1 & \textbf{100\%} & 7.30 & 0.95 & 18\% \\
A-muse-Net - Ablation2 & \textbf{100\%} & 6.80 & 0.90 & 20\% \\
A-muse-Net - Ablation3 & \textbf{100\%} & 7.70 & 0.90 & 18\% \\
A-muse-Net & \textbf{100\%} & \textbf{9.54} & \textbf{0.86} & 29\% \\
\bottomrule
\end{tabular}
\end{center}
\caption{Quantitative Comparison to outher methods.}\label{tab1}
\end{table}

\subsection{User Study}

We asked 17 people to rate the A-Muze-Net generated songs with scores from 1-5 and to state their musical background level, as in \textbf{MuseGan}~\cite{dong2018musegan} and in Lyu et al.~\cite{lyu2020dual}. Each individual listened to ten different clips, which are of length of four-bars, as also was done in the MuseGan research.  Tab.~\ref{tab2} presents their average satisfaction out of our generated songs on a scale from one to five.

As can be seen, our method outperforms the baseline methods. Interestingly, while MuseGAN is second by the user rating, it is far lower on the quantitative results.

\begin{table}
\begin{center}
\begin{tabular}{lc}
\toprule
\textbf{Experiment} & \textbf{US}\\
\midrule
True Music & 3.80\\
Lyu et al. Pianoroll CNN~\cite{lyu2020dual} & 2.40\\
Lyu et al. Embedding Atten-LSTM~\cite{lyu2020dual} & 2.70\\
MuseGan & 3.16 \\
A-muse-Net & \textbf{3.28}\\
\bottomrule
\end{tabular}
\end{center}
\caption{User study results.}\label{tab2}
\end{table}

\section{Conclusions}

While high-capacity models have now been shown to be able to model music based on very large corpora~\cite{dhariwal2020jukebox}, such models remain computationally inaccessible to most research labs and amassing such data if a copy free way is next to impossible. Furthermore, while such models teach us about AI and large-scale pattern extraction, there is little advancement with regard to the foundations of music making.

In this work, we employ a well established machine learning architecture and try to answer fundamental questions about music representation: (1) How to link the melody and the harmony effectively? (2) How to represent symbolic music in an accessible way? (3) How to capture the transient essence of the melody? (4) How to enrich the generated music?

Our answers to each of these questions have led to an improvement over the options that have been used in the previous work. Collectively, our method provides a sizable improvement in all metrics in comparison to the existing methods. 

\section*{Acknowledgements}
This project has received funding from the European Research Council (ERC) under the European Union’s Horizon 2020 research and innovation programme (grant ERC CoG 725974). 

\bibliographystyle{splncs04}
\bibliography{my_bib}

\end{document}